\documentclass[12pt]{article}
\usepackage{epsfig}
\usepackage{latexsym}
\title {Naming Game on Adaptive Weighted Networks}
\author{
 {\bf Dorota Lipowska}\footnote{Corresponding author}\\
Institute of Linguistics, Adam Mickiewicz University, \\
61-874 Pozna\'{n}, Poland\\
e-mail: lipowska@amu.edu.pl, tel: +4861-829-3663\\
{\bf Adam Lipowski}\\
Faculty of Physics, Adam Mickiewicz University, \\
61-614 Pozna\'{n}, Poland\\
e-mail: lipowski@amu.edu.pl, tel: +4861-829-5062\\
}
\date{}
\begin {document}
\maketitle
\begin {abstract}
We examine a naming game on an adaptive weighted network. A weight of connection for a given pair of agents depends on their communication success rate and determines the probability with which the agents communicate. In some cases, depending on the parameters of the model, the preference toward successfully communicating agents is basically negligible and the model behaves similarly to the naming game on a complete graph. In particular, it quickly reaches a single-language state, albeit some details of the dynamics are different from the complete-graph version. In some other cases, the preference toward successfully communicating agents becomes much more relevant and the model gets trapped in a multi-language regime.  In this case gradual coarsening and extinction of languages lead to the emergence of a dominant language, albeit with some other languages still being present. A comparison of distribution of languages in our model and in the human population is discussed. 
\end{abstract}
{\it keywords:} adaptive weighted networks, naming game\\

\newpage
\section{Introduction}
Understanding the evolution of human language is one of the hardest problems in science and remains a great challenge~\cite{CHRISTIANSEN}. Indeed, language is a very complex human behaviour, which results from interactions among a population of individuals, who in turn interact with their environment. In addition, language continuously changes over time by adapting to the environment and the individuals.

A comprehensive approach  to the language evolution should consider it in a wider perspective of communication systems. Trying to identify the most fundamental features of human language, one often lists compositionality, arbitrariness, syntax, cultural transmission, displaced signals, or intentionality~\cite{hockett}. Although still under debate, the list of such features might, nevertheless, be helpful to specify, for example, the differences between  human language and communication systems developed by some other animal species~\cite{parisi}. Moreover, understanding the role of such basic ingredients of language  communication could be vital in constructing artificial organisms (embodied robots or software bots) fitted with  some kind of communication system enabling them to interact with each other or with humans.  There is already a growing number of software agents (programs) collecting and exchanging various information often without any human intervention. An increasing autonomy of such agents might even result in some kind of (spontaneously created) communication system among them and it is certainly desirable that humans should be able to understand and control such a process.  

It is also important to take into account the biological context of language emergence and its subsequent  evolution. Indeed, the emergence of language was accompanied by some genetic changes and is even considered as one of the  major evolutionary transitions of life on Earth \cite{MSMITH1997}. A precise  role of biological factors remains, however, unclear and the attitude of researchers to this issue ranges from nativism \cite{chomsky} to empiricism \cite{sampson} with many scholars taking an intermediate stance.  Further division among researchers is due to the possible adaptive value of language with adaptationists \cite{darwin} and non-adaptationists \cite{gould} at the two extremes. Recent influential paper of Pinker and Bloom strongly advocates an adaptive role of language and has catalysed many linguistic and biological studies~\cite{pinker}.

Closely related to biological considerations are game-theory aspects of communication and language. Sending or receiving information requires some decisions and efforts to be made and some benefits or costs might be related with such actions. However, among individuals competing, for example, for food,  even the very emergence of language is questionable because in such a case deception seems to be the most profitable strategy \cite{dawkins}. The resolution of these dilemmas usually refers to the kin selection~\cite{HAMILTON} or reciprocal altruism~\cite{TRIVERS}. In other words, speakers remain honest because they are helping their relatives or they expect that others will do the same for them in the future. As an alternative explanation Dessalles~\cite{DESSALLES} suggests that honest information is given freely because it is profitable -- it is a way of competing for status within a group.  Some related results on computational modelling of the honest cost-free communication are reported by Noble~\cite{NOBLE}.

From the above exposition it is clear that understanding language emergence and evolution requires considerable efforts \cite{knight,BRIGHTON2005}.
It seems that cultural interactions between users play the major role in shaping the language formation. Recently, a promising approach to model such interactions  is based on computational simulations  \cite{DEBOER,NOWAK,NOWAK2002,PAULO2007,STAUFFER2008}, especially those based on the so-called multi-agent systems~\cite{CANGELOSI,NOLFI} .

An important class of such models originates from the naming game introduced by Steels to describe processes leading to the establishment  of a shared vocabulary, i.e., a set of mappings between words and meanings~\cite{STEELS1995}. In the naming game, agents are involved in pairwise interactions, which direct the model to a state of linguistic synchronization, i.e., to the conformance of agents' vocabularies. Let us emphasize  that while conversations of agents take place strictly locally (each involves only two agents) and go on without any central control, nevertheless the final result of these processes, i.e., a common vocabulary shared by all agents, emerges globally.

To specify the naming game model, one has to define the topology of interactions between agents. For mathematical and computational simplicity, a complete-graph topology is often assumed, where each agent can interact with any other agent~\cite{BARONCHELLI2006}. Another possibility is to place agents at sites of a regular lattice and allow interactions only between the nearest neighbours~\cite{baron2006a}. However, real networks of social interactions seem to be much more complex than the above mentioned graphs~\cite{wasserman,white,barabasi}. To take into account the heterogeneous nature of social interactions, the naming game on scale-free networks and on small-world networks were also studied. Although some quantitative differences in the dynamics were observed, the overall behaviour, namely a relatively fast convergence to the monolingual state, was predominantly the same~\cite{ASTA2006,ASTA2006a,korniss}.

Such a behaviour is in contrast with the multi-language structure of the human population. Indeed, despite gradual extinction, there are thousands of languages still in use, and at a time scale of  tens or hundreds of years, many languages, especially those relatively common ones, seem to be very stable. In the present paper, we show that the naming game defined on an adaptive weighted network allows us to examine such stable multi-language structures. Interactions in our model try to mimic  certain features of social interactions. In particular, the weights of connections between agents differentiate the probabilities of their mutual communication. Moreover, these weights are adaptive and depend on the success rate of past communication attempts of agents. Let us notice that the  formation of multi-language states has already been observed by Dall'Asta {\it et~al.} for complex networks with a community structure~\cite{ASTA2006}. However, their networks are a bit artificial with a community structure imposed "by hand". In our opinion, the multi-language structure of human population is a result of a certain dynamic process, which, we hope, is to some extent captured in our model.

\section{Model and numerical method}
 In the original formulation~\cite{STEELS1995}, the naming game
describes cultural transmission within a single generation
of agents.  Agents are involved in pairwise negotiations and try to establish a common vocabulary for a certain number of objects present in their environment. 
Most of the research, however, is limited to a single-object case (for a recent review see~\cite{loreto-jsm})  since it seems to be sufficient to capture the essence of the dynamics also in a more general case. Multi-object ~\cite{LENAERTS,STEELSINTYRE,LIPLIP2009} as well as evolutionary ~\cite{LIPLIP2008,LIPLIP2008A} versions of the naming game were also studied.

The naming game is also related to models used to describe opinion formation, with the voter model being the prime example~\cite{castellano}. 
Let us stress again that the main emphasis in the naming game model is on the cultural (single-generation) transmission of language. An alternative approach to the language evolution, where inter-generational
interactions play an important role, is called Iterated Learning
Model and was used in some other contexts~\cite{KIRBY2001,STEELS2002}.

In our model we consider a population of $N$ agents playing the single-object naming game.  
Each agent has its lexicon, i.e., a list of words (initially empty). 
Basic steps of the game are as follows:

\begin{enumerate}
\item First, a speaker $i$ and a hearer $j$ are selected ($i \neq j$). Then the speaker selects randomly one of the words from its lexicon (if the lexicon is empty, the agent generates a word randomly)  and communicates this word to the hearer. 
\item The game ends successfully when the hearer has the word in its lexicon; in such a case both agents retain  only the communicated word in their lexicons. 
\item The game fails when the hearer does not have the word in its lexicon;  in this case the word is added to the hearer's lexicon.
\end{enumerate}
After the game, both the number of successes the pair of agents $(i,j)$ have achieved so far as well as the number of all their communication attempts are updated and the communicative success rate for this pair of agents $s_{ij}=s_{ji}$ is calculated as the ratio of these two numbers (of course, initially $s_{ij}=0$ for all pairs of agents).

The above definition corresponds to the so-called minimal version of the naming game~\cite{BARONCHELLI2006}. Of course, the above rules greatly reduce computational complexity, but are to some extent artificial. For example, in the case of success, an immediate obliteration of all the words except the  communicated one  drastically simplifies human memory management.

To complete the definition of our model, we have to specify how we select the speaker and the hearer. Our intention is to simulate an important, in our opinion, criterion: we talk most preferably with those with whom we already have communicated successfully. This is implemented as follows:

\begin{itemize} 
\item The speaker $i$ is selected randomly. 

\item The hearer $j$ is selected using the roulette rule, with the probability 
\begin{equation} 
p_{ij}=\frac{w_{ij}}{\sum_{k = 1}^N w_{ik}}, 
\label{prob}
\end{equation} 

where the weights
\begin{equation} 
w_{ij} = \left\{ 
\begin{array}{ll}
s_{ij}+\epsilon & \textrm{for $i \neq j$} \\
0                   & \textrm{for $i = j$}
\end{array} \right.
\label{weight} 
\end{equation} 

for $i,j=1,2,\ldots, N$.
The (positive and typically small) parameter $\epsilon$ ensures that a speaker can sometimes play the naming game also with such agents with which its up-to-now communicative success rate is very small or even equals zero.
\end{itemize}

Let us notice that the behaviour of  some opinion formation models, which constitute a related class of models, strongly depends on the details of the  selection of the pair of agents. In particular, one can expect that  some properties of our model might change when the hearer is selected prior to the speaker~\cite{nardini}.

From the above rules, it follows that our model constitutes an adaptive weighted network. The weights of links, which depend on the success rates that pairs of agents have achieved so far, change in time and control the intensity of their subsequent interactions. Generally, the bigger the success rate, the more frequent communication attempts, however, due to the parameter $\epsilon$, there is always a possibility to communicate also with such an agent with which there have not been scored any successes so far.

Obviously,  the characteristics of our model change during simulation. To examine their time dependence and compare the results for different numbers of agents~$N$, we have defined a unit of time as $N$ communication attempts (which corresponds to two on average communication attempts per agent). We have monitored a number of time-dependent observables in the system, in particular:

\begin{itemize}
\item  $s$ -- the communicative success rate defined as a fraction of all successes during the last $N$ communication attempts, i.e., within the last unit of time. Let us emphasize that $s$ is calculated for $N$~most recent communication attempts in the entire system of $N$~agents while $s_{ij}$, i.e., the communicative success rate of a pair of agents $i$ and $j$, which determines the weight $w_{ij}$ (\ref{weight}) of the connection between them, stores the entire history of their interactions.
\item $L$ -- the number of different words in lexicons of all agents. At later stages of simulations, when most of agents have only one word in their lexicons, $L$ could be interpreted as the number of languages the agents use at this moment of the evolution of the system.
\item $N_d$ -- the number of agents that have the most common word in their lexicons, i.e., the word that appears in the largest number of lexicons. 
In some cases we measured also the number of users of less common words.
\end{itemize}
To smooth-out statistical fluctuations, the measured quantities $s$, $L$, and $N_d$  
were averaged over independent runs.  Some other observables are described in the following text.

\section{Numerical results}
Initially, all weights $w_{ij}=\epsilon$ for $i\neq j$ and thus the selected speaker chooses any agent as a hearer with equal probability. As the evolution of the model progresses, some pairs of agents  $(i,j)$ might successfully communicate and that increases the corresponding weights~$w_{ij}$. Consequently, an agent will communicate with some agents more often than with others and after some time clusters of such agents will be formed,  with communication taking place mainly within these clusters.
Hence, linguistic synchronization is quickly reached inside clusters and all agents in a cluster eventually use the same language, i.e., they have the same (and only one) word in their lexicons. One should keep in mind, however, that the structure of the network set by the collection of the weights~$w_{ij}$  is dynamic and even strong connections ($w_{ij}\approx 1$) might become weak later on (and also weak connections might get stronger). 

\begin{figure}
\vspace{-3cm} \centerline{ \epsfxsize=13cm \epsfbox{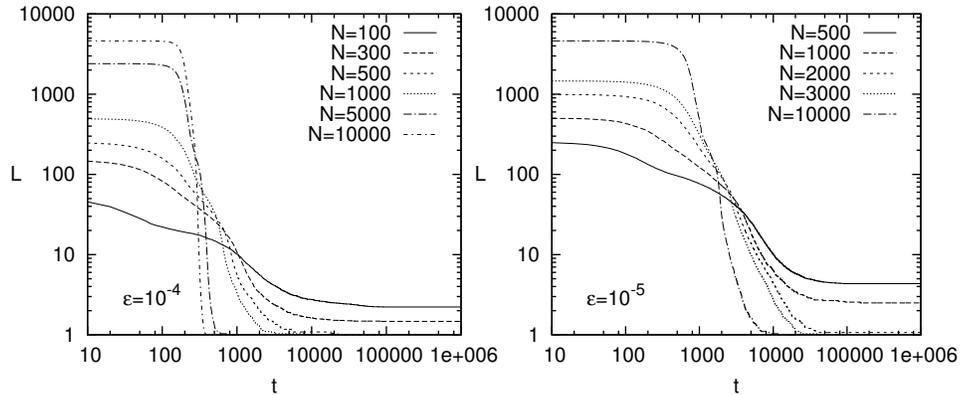} }
\vspace{0cm} \caption{The time dependence of the number of languages $L$ calculated for $\epsilon=10^{-4}$ and $\epsilon=10^{-5}$, and for various numbers of agents~$N$ (logarithmic scales).} \label{1e41e5}
\vspace{-0cm}
\end{figure}

The details of the dynamics and the final state of the model depend on the parameters $N$ and~$\epsilon$. In some cases, the model quickly reaches the regime where all agents have only one and the same word in their lexicons. Such a single-language regime appears in general when $N$ and $\epsilon$ are sufficiently large (Figure~\ref{1e41e5}). Otherwise, dynamics of the model leads to a multi-language regime. In our simulations, except for a relatively short initial interval of simulation time, the number of languages remained nearly constant, which indicates a strong stability of the multi-language regime, similar perhaps to the stability of a glassy phase in supercooled liquids \cite{debenedetti} or in some spin systems \cite{ritort,lipjohn}. 
Although in the limit of large $N$ (and fixed $\epsilon$) the model always enters the single-language regime, for finite $N$ the multi-language phase might exist. 

\subsection{Single-language regime}
Although communication between agents takes place mainly within clusters, some outside-cluster communication attempts are also  made. This is because the weight (\ref{weight}) is positive even for agents with (so far) zero success rate. As a result of such attempts, some agents might change the cluster they belong to;  eventually, some clusters might even disintegrate. Such a process resembles the coarsening dynamics and the order/disorder transition found in other versions of the naming game~\cite{BARONCHELLI2006}.

\begin{figure}
\vspace{-1cm} \centerline{ \epsfxsize=9cm \epsfbox{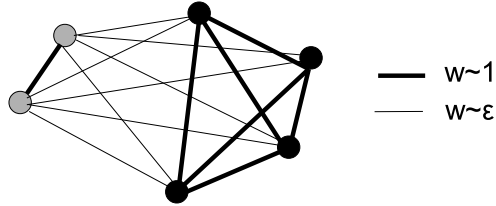} }
\vspace{0cm} \caption{In the final stage of the coarsening process, almost all agents form a single-language cluster (black circles). To include a given (grey) agent which belongs to a certain small cluster into the main cluster, one needs two steps: (i) a black speaker selects a grey hearer, which results in failure, and the grey agent adds to its lexicon the word used by black agents (the rate of this processes scales as $\epsilon N/N=\epsilon$); (ii) the grey agent as a speaker selects a black hearer and the word it chooses to communicate is the one acquired in step~(i) (the rate $\frac{N\epsilon}{N\epsilon+1} \approx N\epsilon$ for small~$N\epsilon$). Because the second step is a success, all words but the communicated one are removed from agents' lexicons, and the grey agent becomes black. The combined rate of steps (i) and (ii) should scale as $N\epsilon^2$. 
} \label{rysunek}
\end{figure}

Obviously, the larger $\epsilon$, the greater the intensity of the outside-cluster communication. The number of agents~$N$ is yet another factor that increases the intensity of such processes. This is because for large~$N$ an agent belonging to a cluster of a given size has so many more candidates to communicate with (albeit of very small weights $\sim \epsilon$).
One might expect that it is the combination of~$\epsilon$ and~$N$, rather than those parameters taken separately, which is the control parameter of the model.  Although we cannot provide a rigorous derivation, some arguments (Figure~\ref{rysunek}) suggest that in our model, some important processes, which most likely underlie the coarsening,  take place at a rate that scales as~$N\epsilon^2$. The numerical results presented below are collected in sets corresponding to constant $N\epsilon^2$ and their reasonably good convergence (for increasing~$N$) seems to support the suggestion that $N\epsilon^2$ indeed is the control parameter of the model. 

\begin{figure}
\vspace{0cm} \centerline{ \epsfxsize=9cm \epsfbox{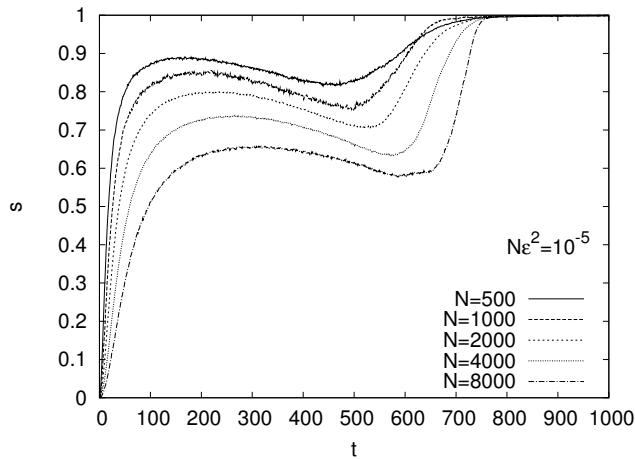} }
\vspace{0cm} \caption{The time dependence of the success rate $s$ calculated for several values of $N$ and for $N\epsilon^2=10^{-5}$.} 
\label{sukces-mono}
\end{figure}

Our simulations show that when $N\epsilon^2$ is sufficiently large, the outside-cluster communication is sufficiently frequent and the  model behaves similarly to the naming game on a complete graph, where it is known to reach a state of complete synchronization~\cite{BARONCHELLI2006}. Our numerical results confirm basically such a behaviour but exhibit some important differences as well. In Figure~\ref{sukces-mono} we present the success rate~$s$ calculated for $N\epsilon^2=10^{-5}$. One can notice that after an initial interval, agents reach the state where they communicate with a large success rate $s\sim 1$. Interestingly, this initial interval is almost the same ($t\sim 750$) for all examined systems even though their sizes~$N$ differ substantially.

\begin{figure}
\vspace{0cm} \centerline{ \epsfxsize=9cm \epsfbox{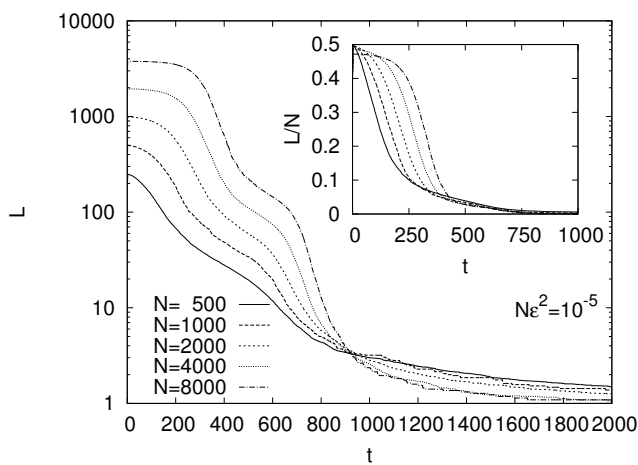} }
\vspace{0cm} \caption{The time dependence of the number of languages $L$ (logarithmic scale) calculated for several values of $N$ and for $N\epsilon^2=10^{-5}$. The inset shows the time dependence of the normalized number of languages $L/N$.} 
\label{jezyki-mono}
\end{figure}

\begin{figure}
\vspace{-1cm} \centerline{ \epsfxsize=9cm \epsfbox{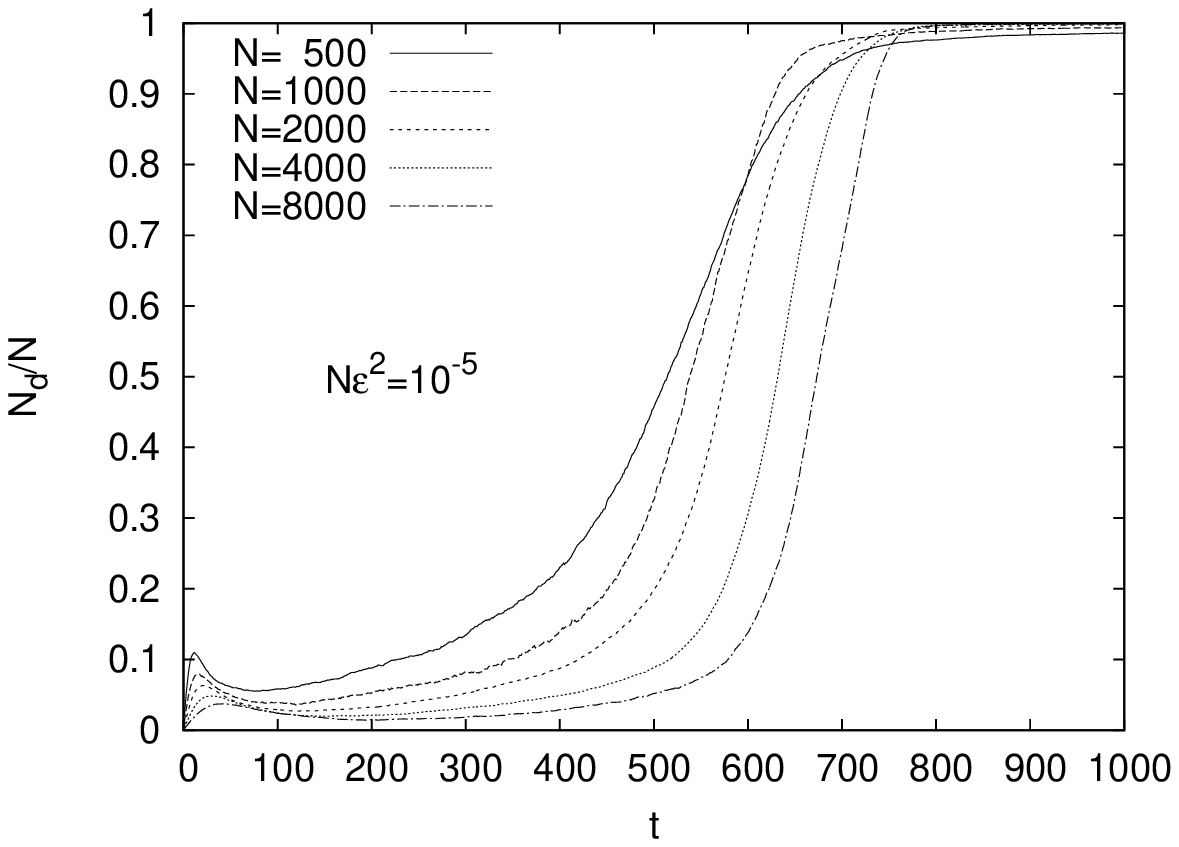} }
\vspace{0cm} \caption{The time dependence of the ratio of agents that use the most common language $N_d/N$ calculated for several values of $N$ and for \mbox{$N\epsilon^2=10^{-5}$}.} 
\label{domin-mono}
\end{figure}

Since the number of languages $L$ diminishes in time  (Figure~\ref{jezyki-mono}), we conclude that indeed the system evolves toward the single-language state. The inset of Figure~\ref{jezyki-mono} shows that at the same time when the success rate~$s$ approaches unity ($t\sim 750$, Figure~\ref{sukces-mono}), the normalized number of languages $L/N$ drops to 0. Let us also note that the crossing point around $t=900$ in Figure~\ref{jezyki-mono} suggests that in the limit $N\rightarrow\infty$ for $t>900$, there is only one language in the system. Figure~\ref{domin-mono}, where the ratio of users of the most common language $N_d/N$ is plotted, shows that for $t>750$ almost all agents use the same language. 

\begin{figure}
\vspace{-4cm} \centerline{ \epsfxsize=13cm \epsfbox{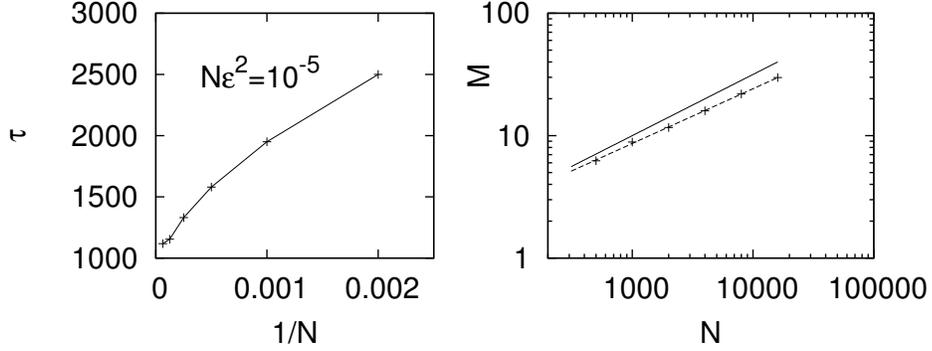} }
\vspace{0cm} \caption{Left panel shows the size dependence of the average time needed to reach a single-language state. Calculations were done for $N=500$,\ 1000,\ 2000,\ 4000,\ 8000,\ 16000 and for $N\epsilon^2=10^{-5}$. For each~$N$, usually 100 independent runs were made. Right panel shows the size dependence of the averaged maximum number of words $M$ in agents' lexicons ($N\epsilon^2=10^{-5}$). The fitted dashed line has the slope~0.45 and the continuous line corresponds to the complete graph (slope 0.5).} 
\label{time}
\end{figure}

As it is already known, a characteristic time $\tau$  to reach a single-language state in the naming game increases with the system size~$N$. For example, it is found that on a complete graph $\tau\sim N^{1.5}$~\cite{BARONCHELLI2006}.
To compare this result with the behaviour of our model, we calculated $\tau$ for several values of $N$ and the results are presented in Figure~\ref{time}. They show that $\tau$ does not diverge for increasing~$N$ but most likely  converges to a finite value~$\approx 1050$, which is approximately consistent with the time scale seen in the behaviour of $s$, $N_d/N$, and $L$. An additional comparison with a complete-graph version can be obtained from the analysis of the number of words kept in the lexicons of agents. Typically, this number (as a function of time~$t$) first gradually increases and then decreases as the single-language state is being approached. It is known that the averaged maximum number of words (per agent)~$M$ on a complete graph increases asymptotically as $N^{0.5}$~\cite{BARONCHELLI2006}. Numerical results for our model, shown in the right panel of Figure~\ref{time}, suggest that $M$  increases asymptotically as $N^{0.45}$. Such a discrepancy is yet another indication that the dynamics of our model, despite qualitative similarities, differs from the complete-graph version.

\subsection{Multi-language regime}

\begin{figure}
\vspace{-2cm} \centerline{ \epsfxsize=13cm \epsfbox{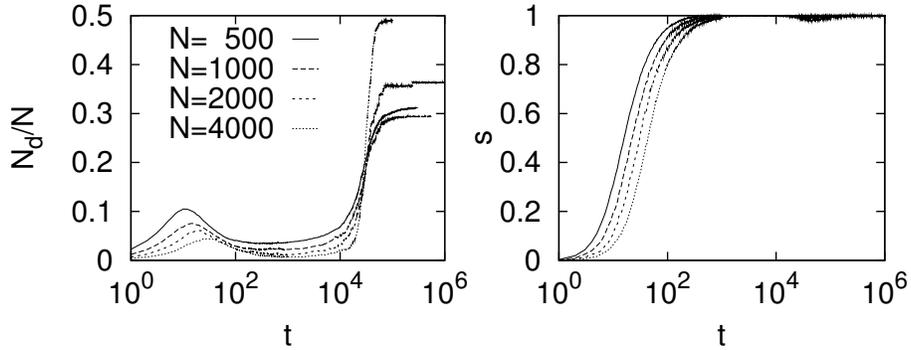} }
\vspace{0cm} \caption{Left panel shows the time dependence of the ratio of agents that use the most common language $N_d/N$ calculated for several values of $N$ and for \mbox{$N\epsilon^2=10^{-9}$}. The right panel shows the time dependence of the success rate $s$ for the same values of~$N$ and $N\epsilon^2$.} 
\label{domin-poly}
\end{figure}

\begin{figure}
\vspace{0cm} \centerline{ \epsfxsize=9cm \epsfbox{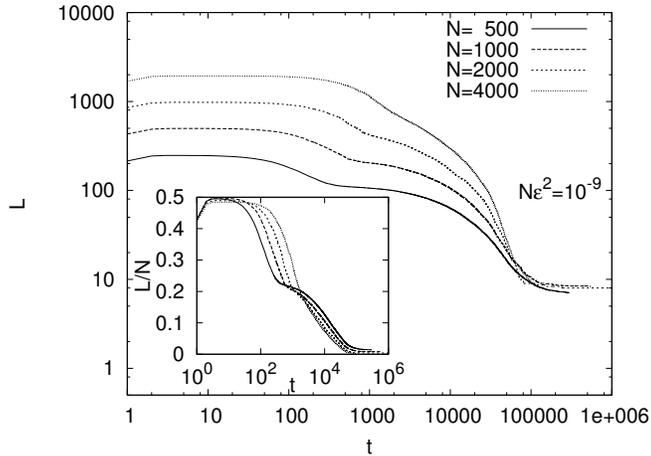} }
\vspace{0cm} \caption{The time dependence of the number of languages $L$ calculated for several values of $N$ and for $N\epsilon^2=10^{-9}$. The inset shows the time dependence of the normalized number of users $L/N$.} 
\label{jezyki-poly}
\end{figure}

When  $N\epsilon^2$ is sufficiently small, the outside-cluster communication attempts are very rare, which results in a much different behaviour of our model. In the right panel of Figure~\ref{domin-poly}, one can notice that the success rate~$s$ reaches unity around $t=10^3$, however, at that time the ratio of  agents using the most common language $N_d/N$ is much smaller than unity.
Only after \mbox{$t\approx 3\cdot 10^4$}, this fraction significantly increases, which suggests the emergence of a dominant language. However, the dominant language only partially invades the system, since some other languages also persist, apparently for an arbitrarily long time (Figure~\ref{jezyki-poly}). Indeed, around $t=10^5$ the number of languages saturates at $L \sim 10$ and does not seem to diminish even up to $t=10^6$. Although there are several languages for $t>10^5$, their number is only a small fraction of the number of agents~$N$ (inset in Figure~\ref{jezyki-poly}).

In the multi-language regime one can thus distinguish three phases. The first one ($t<10^3$) can be called pre-linguistic since the success rate remains relatively low. Next ($10^3<t<3\cdot 10^4$) there is a phase with many languages but without a dominant language ($N_d/N\ll 1$). In this phase the number of languages gradually decreases. In the third phase ($t>3\cdot 10^4$) several languages exist but one of them emerges as a dominant one and is used by a large fraction of agents. Let us notice that for increasing~$N$, the emergence of a dominant language becomes more and more abrupt.

\begin{figure}
\vspace{0cm} \centerline{ \epsfxsize=9cm \epsfbox{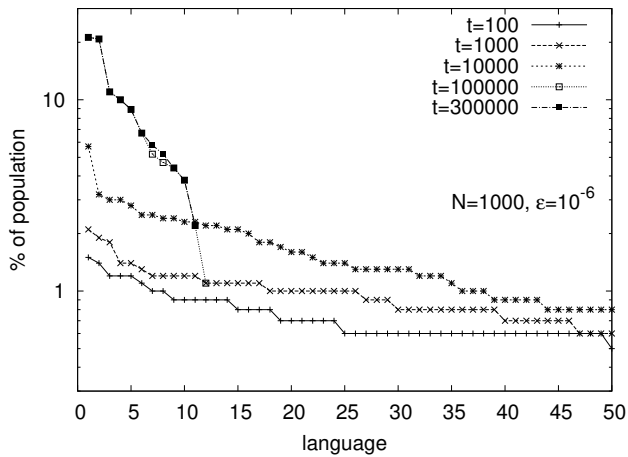} }
\vspace{0cm} \caption{The percentage of population constituted by the users of the  50 most common languages, calculated for  $N=1000$ and $\epsilon=10^{-6}$ (i.e., in a multi-language regime).} 
\label{dist1e6}
\end{figure}

\begin{figure}
\vspace{0cm} \centerline{ \epsfxsize=9cm \epsfbox{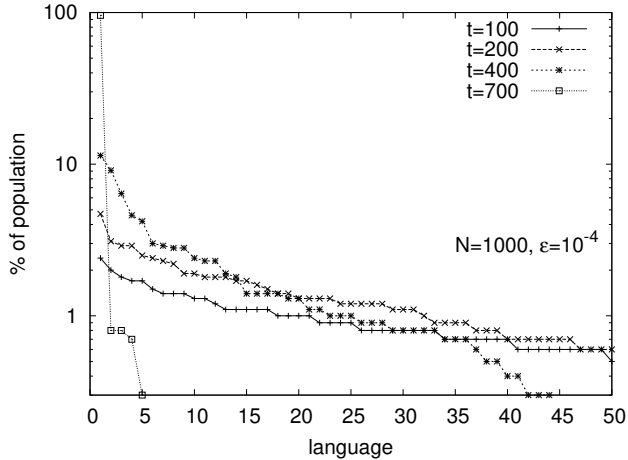} }
\vspace{0cm} \caption{The percentage of population constituted by the users of the  50 most common languages, calculated for  $N=1000$ and $\epsilon=10^{-4}$ (i.e., in a single-language regime).} 
\label{dist1e4}
\end{figure}

To get some insight into a possible mechanism that stabilizes the multi-language state, we calculated the numbers of users of the 50 most common languages. We made only a single run (no averaging over independent runs) and Figure~\ref{dist1e6} shows how these numbers change in time. Initially, one can observe a gradual increase in the numbers of users of the most common languages at the expense of the  less prevalent ones. Around $t=10^5$ only 12 languages are left and at $t=3\cdot 10^5$ the least common one (among these 12) becomes extinct. One can notice that there are virtually no other changes between the data at $t=10^5$ and $t=3\cdot 10^5$. Let us also observe that the remaining languages are relatively prevalent and even the least frequently used one has more than 20 users (2\% of the population).  For comparison, we present analogous data from the single-language regime (Figure~\ref{dist1e4}). Although transiently a number of relatively common languages are formed in this case, eventually only one of them survives. Thus it is not a sheer number of users that determines whether a language will survive or not but also the value of the control parameter~$N\epsilon^2$, which plays an important role.


\begin{figure}
\vspace{0cm} \centerline{ \epsfxsize=9cm \epsfbox{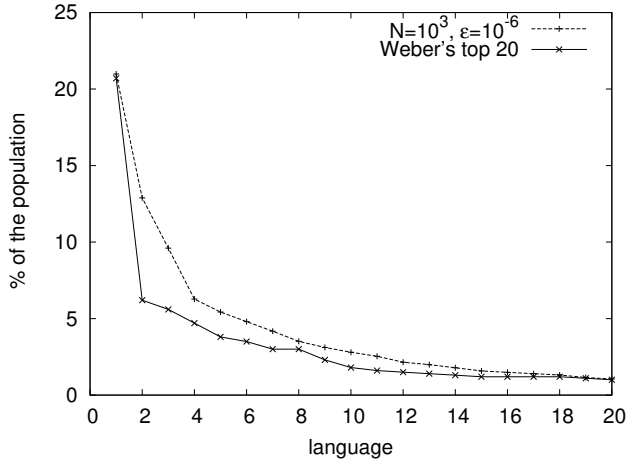} }
\vspace{0cm} \caption{The percentage of population constituted by the users of the  20 most common languages, calculated for  $N=1000$ and $\epsilon=10^{-6}$ (i.e., in a multi-language regime) compared with Weber's statistical data~\cite{weber1997}. The first three languages according to Weber are Chinese (20.7\% of the population, i.e., $1.1\cdot 10^9$ speakers), English (6.2\%, $3.2\cdot 10^8$), and Spanish (5.6\%, $3.0\cdot 10^8$).} 
\label{weber}
\end{figure}

We are tempted to think that the behaviour of our model in the multi-language regime bears  some similarity to the evolution of human languages.  If so, some very basic characteristics of human languages  could be reproduced within our model. In Figure~\ref{weber}, we present the distribution of users of the 20 most common existing languages~\cite{weber1997} compared with analogous distributions obtained for our model. We made a simulation for $N=10^3$ and $\epsilon=10^{-6}$, which corresponds to the multi-language regime.
The simulation time was chosen in such a way that the percentage of users of the most common language was equal to 20.7\%, i.e., the percentage of speakers of Chinese.
Let us notice that there is a reasonably good agreement even for less common languages and perhaps by modifying the value of $N\epsilon^2$ (within a multi-language regime), a still better fit could be achieved. However, taking into account the complexity of the evolution of human language, such an agreement is most likely accidental and should be considered with great care.  For example, such important factors responsible for the distribution of languages as population changes due to births and deaths are entirely neglected in our approach. Let us notice, however, that there are some naming game models implementing such processes \cite{LIPLIP2008}. Partially, the problem of populational factors could be taken into account considering only languages of populations with a similar growth rate (for example only European languages) but further analysis along this line is left for the future.

\section{Remarks and Conclusions}
In the present paper we examined the naming game on an adaptive weighted network. In some cases, depending on the parameters of the model, the preference due to successful communication in the past is basically negligible and the  model behaves similarly to the naming game on a complete graph. In particular, it quickly reaches the state of complete synchronization where all agents use the same language. However, the average time needed to reach such a state does not diverge as the number of agents~$N$ increases --- contrary to the naming game on complete graphs. Moreover, the average maximum number of words in the lexicons of agents increases as $N^{0.45}$, which is close but different than the $N^{1/2}$ increase expected for the naming game on a complete graph.

For some other values of the parameters, the preference due to  successful communication in the past is much more relevant and quite different and perhaps more interesting behaviour of our model appears. In this case our model does not reach complete synchronization but remains trapped in a multi-language state. Three phases can be distinguished in the evolution of our model. In the first one, the average success rate of communicating agents is rather low and most likely this phase corresponds to a prelinguistic phase. In the next phase, although the average success rate is close to unity, many languages exist in the system and none of them dominates. At a certain moment, however, the model enters the third phase where many languages still exist but one of them becomes dominant. Since languages are dynamically equivalent, the emergence of the dominant language might be considered as some kind of spontaneous symmetry breaking. Interesting issues of broken ergodicity or stability of the resulting structures are, however, omitted in the present paper.

We also would like to mention that reaching the consensus in naming-game models shows some similarity to analogous processes in opinion formation models. Trapping the dynamics in a multi-language state in our model resembles cluster formation in some opinion-formation models where some adaptive rewiring mechanisms generate strong community structure \cite{gil,holme,kozma}. It would be interesting to examine further relations with these two classes of models. 

One can speculate about the possibility that human languages evolved similarly to the multi-language scenario. Such an idea gets  some support from the comparison of Weber's distribution of the 20 most common languages and the analogous distribution in our model.
An interesting, in our opinion, question is whether the rapidly progressing extinction of less widespread ethnic languages will lead to the emergence of a dominant language and if so, when such a situation occurs. With this respect, our work  is, of course, inconclusive. Let us notice, however, that the users of the most common language currently constitute about~$20\%$ of the human population and according to Figure~\ref{domin-poly}, we might be very close to the transition where the dominant language will emerge.  We should emphasize, however, that our model is very simple and its applicability to real linguistic data should be considered with care. For example, a constant number of agents $N$ is not consistent with fluctuations in real human populations due to, e.g., demographic processes. Perhaps it would be more appropriate to compare our numerical
data rather with the distribution of dialects of a sufficiently widespread language such as, for example, Chinese.

For models on adaptive networks, there is a subtle point concerning the time scales that govern the evolution of the network itself and of the model (in our case, the naming game). In principle, these processes might have two independent time scales, which in some cases might even be well separated~\cite{gross-sayama,blasius}. In our model, however, the evolution of weights is strongly coupled with  agreeing dynamics of the naming game. Since the success rate stores the entire history of communication between agents, the weights  (\ref{weight}) typically lag behind the agreeing dynamics.  In particular, these weights will evolve even when the naming game has already reached the single-language state (in such a case, the weights will gradually evolve toward $w_{ij}=1+\epsilon$ for every $i\neq j$).
It would be desirable to analyse the interplay of these two time scales in more detail, for example, in the case when the success rate stores only a limited number of last communication attempts. It would also  be  interesting to have a better understanding of the strong stability and long lifetimes of languages in the multi-language regime.  One can consider multi-language states as metastable states of the dynamics and perhaps there is some analogy with metastable states in a certain voter model   on complex networks with a strong community structure~\cite{castello}. We also suggested some similarity of the multi-language state with the glassy state of some physical systems. Further studies on this subject as well as a more detailed analysis of the structure of the adaptive network is, however, left for the future.

\textbf{Acknowledgments:} D.L. is supported with NCN grant 2011/01/B/HS2/01293. We gratefully acknowledge access to the computing facilities at Pozna\'n Supercomputing and Networking Center. 


\end {document}